# Motional resonances of three-dimensional dual-species Coulomb crystals

**Byoung-moo Ann[1], Fabian Schmid , Jonas Krause, Theodor W Hänsch, Thomas Udem  and Akira Ozawa**

Max-Planck-Institut für Quantenoptik, Hans-Kopfermann-Straße 1, D-85748 Garching, Germany

E-mail: b.ann@tudelft.nl and akira.ozawa@mpq.mpg.de

## Abstract

We investigate the motional resonances of dual-species Coulomb crystals comprised of $^9Be^+$ and $^{24}Mg^+$ ions held in a 4-rod linear Paul trap. Our experimental data and simulations show that the secular motion of such mixed crystals has rich dynamics. Their secular spectra can differ significantly from those of pure ion crystals. We propose a simple model based on mechanical coupling with Coulomb interactions between the two different ion species that explains many features of the secular spectrum. Our findings contribute to a more reliable identification of the ion species in mixed crystals.

## 1. Introduction

Laser cooling of atoms [1] and trapped ions [2] has allowed to suppress systematic effects such as Doppler broadening and shifts and has set the stage for today's optical precision metrology. Laser cooling techniques nowadays even make it possible to cool down to $\mu K$ or nK temperatures [3–5] and in some cases to the motional ground state [6, 7]. This development was accompanied with the emergence of highly stable laser systems [8–10] to probe narrow-band transitions. Sympathetic cooling provides a way to cool molecular or atomic ion species that do not possess suitable cooling transitions [11–20]. In this scheme the sympathetically cooled ions are mixed with coolant ions which have a convenient cooling transition. Coulomb interaction between the ions then leads to thermalization of the whole ensemble. With sufficient cooling power, the two ion species eventually form mixed Coulomb crystal structures that have been described in previous publications [11–17, 21–23].

While the coolant ions can be easily located by observing their fluorescence induced by the cooling laser, it is more difficult to detect the presence of the sympathetically cooled ions. Observation of dark spots within the Coulomb crystal is one method, but this is not very quantitative in large crystals, and offers little or no information on the type of ion. Exciting the secular motional resonances is another widespread method [16, 24, 25]. In this case the motion is coupled to the coolant ions which can be observed optically [26, 27]. The secular motion leads to Doppler broadening of the cooling transition which typically results in an increased or decreased fluorescence rate when the cooling laser is far red detuned or near resonance respectively.

In contrast to a large Coulomb crystal, the secular motion of a single trapped and laser cooled ion is simple and can be described as a damped harmonic oscillator. The strength of the confinement is given by the trap voltages and the charge-to-mass ratio of the ion. The confinement force, the ion mass, and the dissipative force induced by the cooling laser determine the frequency and the quality of the resonance. Once more than one ion form a Coulomb crystal, their mutual repulsion gives rise to additional forces. Therefore, their motional resonances are different from those of the individual ions[2]. It is of importance to model such mechanical coupling between the ion species to better understand the secular spectra of mixed ion crystals. The longitudinal vibration modes of a linear chain of identical ions can be readily obtained, with up to 3 ions even in analytical form [28]. This model may be readily extended to mixed ions and radial vibrations. Experimentally, a linear chain is obtained with not too many ions stored in a linear Paul trap and/or with strong radial confinement. Mixed crystals that consist of a few ions find important applications in quantum information processing and precision spectroscopy [13, 29, 30]. When the number of ions exceeds a certain threshold or the axial confinement is strong, a three-dimensional structure is formed [31]. Mixing two species in this case leads to a separation of the species [11, 12]. Ions with a smaller mass-to-charge ratio experience stronger confinement and form a 'core' surrounded by a 'shell' of heavier ions. In the case of a three-dimensional structure consisting of a large number of ions, it is no longer feasible to obtain exact analytical solutions for the secular spectra. A simple explanation of the mechanical coupling between the two ion species is given by Roth *et al* [16] and Zhang *et al* [17, 22]. In this model, the potential induced by the core or shell ions significantly changes the potential of the other species. The dynamics of sympathetically cooled ions under the influence of the space charge of laser cooled ions are also modeled by Baba and Waki [25].

[1] Present Address: Kavli Institute of Nanoscience, Delft University of Technology, Lorentzweg 1, 2628 CJ Delft, The Netherlands.

[2] One exception is the center-of-mass motion of the ion crystal, where the total mass-to-charge ratio still uniquely determines the secular frequency.

Even though it has been shown that the secular motion of mixed ion crystals behaves differently from that of pure ion crystals, there was still no full understanding how the two ion species mechanically interact and lead to a modified secular spectrum. In this work, we present an intuitive model that describes the coupling between two ion species due to the Coulomb interaction. Based on this model, we find that the secular spectra of ion crystals with a 'core' and 'shell' structure exhibit characteristic features over a wide range of crystal parameters. The resonance of the lighter ion species is broadened and shifted to higher frequencies, while the resonance of the heavier species is split into two components. Our model, which partially is an extension of the space charge model proposed in [22], explains the origin of these effects. The model is confirmed both by experiment and by MD simulation, where the secular spectra of the individual ions are investigated. Most experimental studies of two-species trapped ion crystals have been performed in the limit of very few ions [13, 29, 30], or in situations where only one of the species (the coolant) can be observed directly via laser induced fluorescence imaging [14–16, 19, 23]. Here, we report on experiments where both ion species, Be$^+$ and Mg$^+$, permit direct laser imaging, so that experiments and molecular dynamics simulations can be readily compared. The paper is organized as follows. In section 2 we study the motional properties of the mixed ion crystals by employing MD simulations. Several interesting features are highlighted, and a simple model is presented to explain them. Experimental results are given in section 3. The last section summarizes our findings.

## 2. Theoretical description

### 2.1. Mixed Coulomb crystal

MD simulations are widely used to investigate the motion of atoms and molecules [32]. We use this method to model the dynamics of our mixed Be$^+$ and Mg$^+$ Coulomb crystals. We developed an MD software written in C++ which uses the fifth-order Runge–Kutta method with adaptive time steps for solving the multi-particle equations of motion.

As sketched in figure 1(a), our trap is a 4-rod linear Paul trap with ring electrodes for axial confinement [33]. To save computation time, in the MD simulation, the time-averaged secular potential is employed

$$\Phi(x, y, z) = \frac{e^2 V^2}{4m\Omega^2 r_0^4}(x^2 + y^2) + \frac{\kappa e U}{z_0^2}\left(z^2 - \frac{1}{2}(x^2 + y^2)\right).$$
(1)

Here, $V = 417.2$ V and $\Omega = 2\pi \times 36.5$ MHz are the amplitude and the radio frequency (RF) applied to the rod electrodes, while $U = 265$ V is the dc voltage applied to the ring electrodes. The distance between the two ring electrodes is $2z_0 = 1.85$ cm with $\kappa = 0.021$ being a geometrical factor that depends on the diameter of the rings. The distance of the rod electrodes to the trap axis is given by $r_0 = 1.475$ mm. The values assigned here are the ones used in the experimental setup and in the simulations. The time-averaged potential given in equation (1) ignores possible effects due to the micromotion. The approximation is justified because the time scale of the secular motion is an order of magnitude slower than that of the micromotion. This is confirmed by comparing the results of some simulations using the full time-dependent potential with results obtained using the time-averaged secular potential (see appendix A).

To model the laser cooling, we assume an isotropic continuous drag force $\vec{f}_{drag} = -\alpha\vec{v}$ that acts on both ion species. In the experiments we also laser cool both ion species. The linear dependence on velocity $\vec{v}$ is a good approximation when harmonic oscillations at the secular frequencies $\omega_j$ ($j = x, y, z$) lead to Doppler shifts that are much smaller than the line width of the cooling transition. The drag force model also assumes the weak binding regime, that is, the secular frequencies are much smaller than the line width of the cooling transition. In this regime there are many absorption and emission events per secular oscillation period such that a continuous drag force is a good approximation. This drag force leads to a vanishing temperature. In a real experiment random momentum kicks due to spontaneous emission and collisions with the background gas would give rise to heating effects and lead to a finite temperature even when laser cooling is engaged. We have run some simulations that include heating [17] and find that a finite temperature of a few mK does not affect our qualitative description of motional coupling of mixed ion crystals. Hence, we ignore these heating effects for the results obtained in this work.

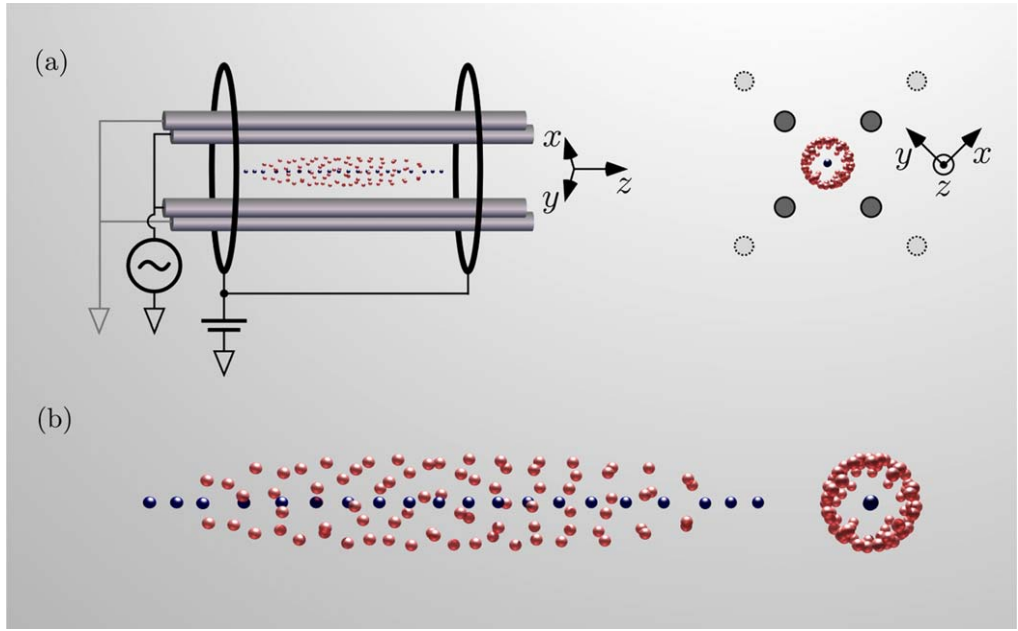

**Figure 1.** (a) Ion trap and geometry used in this work, radial (left) and axial view (right). The cooling and photo-ionization beams for each ion species are directed along the trap axis (*z*). The ion crystals are imaged on an EMCCD camera (not shown) looking perpendicular on the trap axis. The outer electrodes depicted with dashed circles (axial view only) are for compensation of stray fields. An alternating voltage is applied to one of the outer electrodes in order to excite secular motion of the ions. (b) A dual-species Coulomb crystal obtained by the molecular dynamics simulation. It consists of 80 $Mg^+$ ions (light red spheres) and 20 $Be^+$ ions (dark blue spheres). Left: radial view; right: axial view of the same Coulomb crystal.

As can be seen from equation (1), the radial trapping potential is inversely proportional to the mass of the ions *m*. Therefore, lighter ions are more strongly bound to the trap center than heavier ions. Secular frequencies for a single $^{24}Mg^+$ ion are approximately $\omega_z = 2\pi \times 115$ kHz along the axial direction and $\omega_{x,y} = 2\pi \times 370$ kHz along the radial direction. For a single $^9Be^+$ ion the frequencies are $\omega_z = 2\pi \times 187$ kHz and $\omega_{x,y} = 2\pi \times 1.0$ MHz. Under this potential, $^{24}Mg^+$ experiences a radial restoring force of $2.1 \times 10^{-19}$ N at a distance of 1 $\mu$m from the trap axis. The drag force constant $\alpha$ was set for both ion species to $2.58 \times 10^{-21}$ kg s$^{-1}$ in our simulations. This value affects the cooling time and the spectral width of the vibrational resonances, but has little influence on the secular frequencies.

In the simulations the ions' initial positions are randomly chosen, and their initial velocities are set to zero. In the next step the ions are subjected to the trapping and cooling forces until their positions reach a steady state. For the parameters given in this work, 3 ms is enough to crystallize about one hundred ions. An example of a crystal structure obtained in this way is given in figure 1(b). As expected, the $Mg^+$ ions form a shell structure surrounding the $Be^+$ ions that are located along the trap center.

### 2.2. Secular excitation

Once the ions are crystallized by the trap potential and the drag force, we introduce a sinusoidal excitation force along the *x* direction $\vec{f}_{ex} = \hat{x} f_0 \cos(\omega t)$ to induce motional excitation. Here, $\hat{x}$ is a unit vector along the *x* direction, $f_0$ is the amplitude of the excitation force, and $\omega/2\pi$ is the excitation frequency. In the experiment, the force is induced by modulating the voltage of one of the outer electrodes (see figure 1(a)). Since the distance between the electrode and the trap center is sufficiently large compared to the size of the ion crystal, a uniform force is a good approximation. In principle the ion motion is not exactly harmonic, but contains non-linearities due to the Coulomb interaction between the ions. These nonlinearities become negligible when the driving force is weak enough, such that the ion positions do not deviate too much from their equilibrium positions. In this work, we focus on the fundamental frequencies and therefore employ only small amplitudes of the driving force by setting $f_0 = 10^{-20}$ N in the MD simulation. This is much smaller than the typical radial restoring force of the trap. As a result, the non-harmonic component of the ion motion is negligible in the simulation.

After the excitation force is turned on at $t = 0$, the ions start oscillating with the frequency $\omega$. The oscillation amplitude first grows and after some time (approximately 300 $\mu$s in our simulations) stabilizes by balancing with the cooling force. In this state we determine the amplitude $\text{Max}(x_i(t)) - \text{Min}(x_i(t))$ and the mean position $\langle x_i \rangle$ averaged over one oscillation period. The ions are numbered by the index *i*. We only consider the motion along the *x* direction since the motions along *y* and *z* are much smaller (about 10% of the motion along *x* direction). From this data we determine the secular spectrum as the amplitude averaged over all $Mg^+$ or all $Be^+$ ions. An example is given in figure 2 together with the secular spectra of a single $Be^+$ ion and a single $Mg^+$ ion for comparison. The $Be^+$ spectrum in the mixed crystal is significantly broadened compared with the single ion spectrum and the peak is found to be up-shifted (by 18.5 kHz). A more dramatic change is observed for $Mg^+$. Its secular resonance is split into two components, one of which is down-shifted to 276 kHz, and the other is up-shifted to 408 kHz compared with the single ion resonance at 370 kHz.

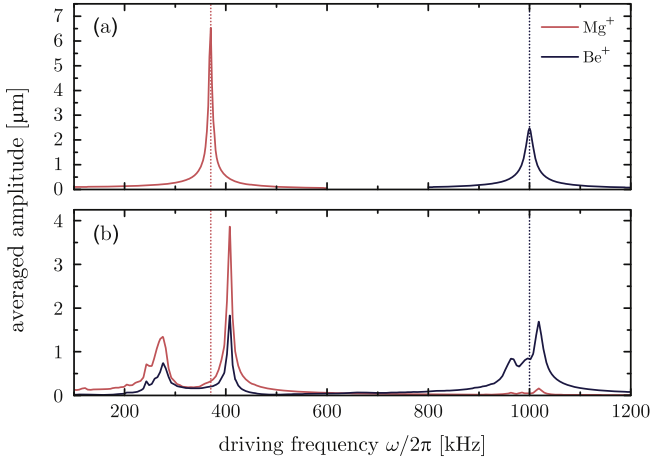

**Figure 2.** The secular spectrum calculated by the molecular dynamics simulation. (a) The light red and the dark blue curves refer to the secular spectrum of a single trapped $Be^+$ ion and a single trapped $Mg^+$ ion respectively, centered at the vertical dashed lines. (b) Secular spectrum of all $Be^+$ and all $Mg^+$ ions combined within the mixed Coulomb crystal shown in figure 1(b). The motion of the two species is coupled due to the mutual Coulomb interactions. Therefore, the resonances of the $Mg^+$ ions also show up in the spectrum of the $Be^+$ ions and vice versa.

### 2.3. Secular spectrum: $Be^+$ core ions

To obtain a more detailed picture, we investigate the secular spectra of individual ions within the crystal. We first have a look at the 20 $Be^+$ core ions of the crystal in figure 1(b). Apparently all these spectra are different as shown in figure 3. For the case of the core ions located near the center, the secular resonance tends to be up-shifted, while the ions located at the ends experience a down-shift of their secular resonance. These effects yield an inhomogeneous broadening of the averaged spectrum as shown in the bottom plot of figure 3. A qualitative explanation of the position dependence can be given in following way: the $Mg^+$ ion shell generates a potential in addition to the trap potential[3]. For the core ions in the center of the trap, this leads to an enhancement of the confinement and hence to a larger secular frequency. On the other hand, the core ions at the ends are not surrounded by the repelling shell ions and hence experience a de-confinement and therefore a reduction of the secular frequency. Qualitatively this behavior can be seen in figure 3, even though this simple picture neglects the motion of the shell ions while the motion of the core ions is excited. This is a reasonable approximation as can be seen from figure 2 for an excitation close to the single $Be^+$ ion secular frequency (1 MHz).

---

[3] One might think that the ion shell does not generate a potential well on its inside. This, however, is only true for an infinitely long cylinder with homogeneous and symmetric charge distribution.

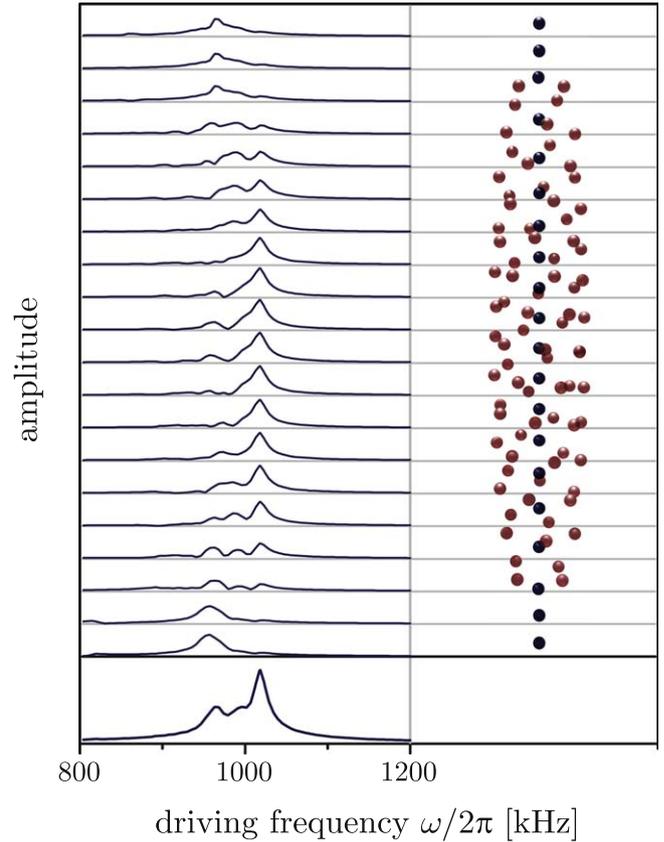

**Figure 3.** Secular spectra of the individual $Be^+$ (core) ions (in dark blue) in the mixed Coulomb crystal of figure 1(b). The bottom plot shows the average that is also seen in figure 2(b).

One noticeable feature is that the secular spectra of individual $Be^+$ ions having symmetric positions with respect to the trap center are not identical. The reason for these differences is that the ion crystal is not exactly symmetric when flipping the $z$-axis. Depending on the exact configuration of their surroundings, ions at symmetric positions experience different local potentials. It should be noted that the asymmetry is not due to the random initial conditions of the simulation. The most stable ion configuration does not necessarily exhibit mirror symmetry with respect to the $xy$-plane.

### 2.4. Secular spectrum: $Mg^+$ shell ions

Next we investigate the motion of the 80 $Mg^+$ shell ions of the ion crystal in figure 1(b) in order to understand the resonance splitting near the single $Mg^+$ resonance into two peaks around 276 kHz and around 408 kHz (see figure 2(b)). The excursions along the $x$-axis of these ions in the time domain are shown in figures 4(a) and (b) for these two frequencies as obtained from our MD simulation. At 276 kHz it appears that ions with smaller magnitude of $\langle x_i \rangle$ experience a larger excursion than those with larger magnitude of $\langle x_i \rangle$. For the peak at 408 kHz it is the other way around. The situation can be clearly seen in figures 4(c) and (d), where the correlation between the motional amplitude and averaged position $\langle x_i \rangle$ is shown. Figures 4(e) and (f) sketch the motion of the shell ions in the axial view represented by the lengths of the green horizontal arrows.

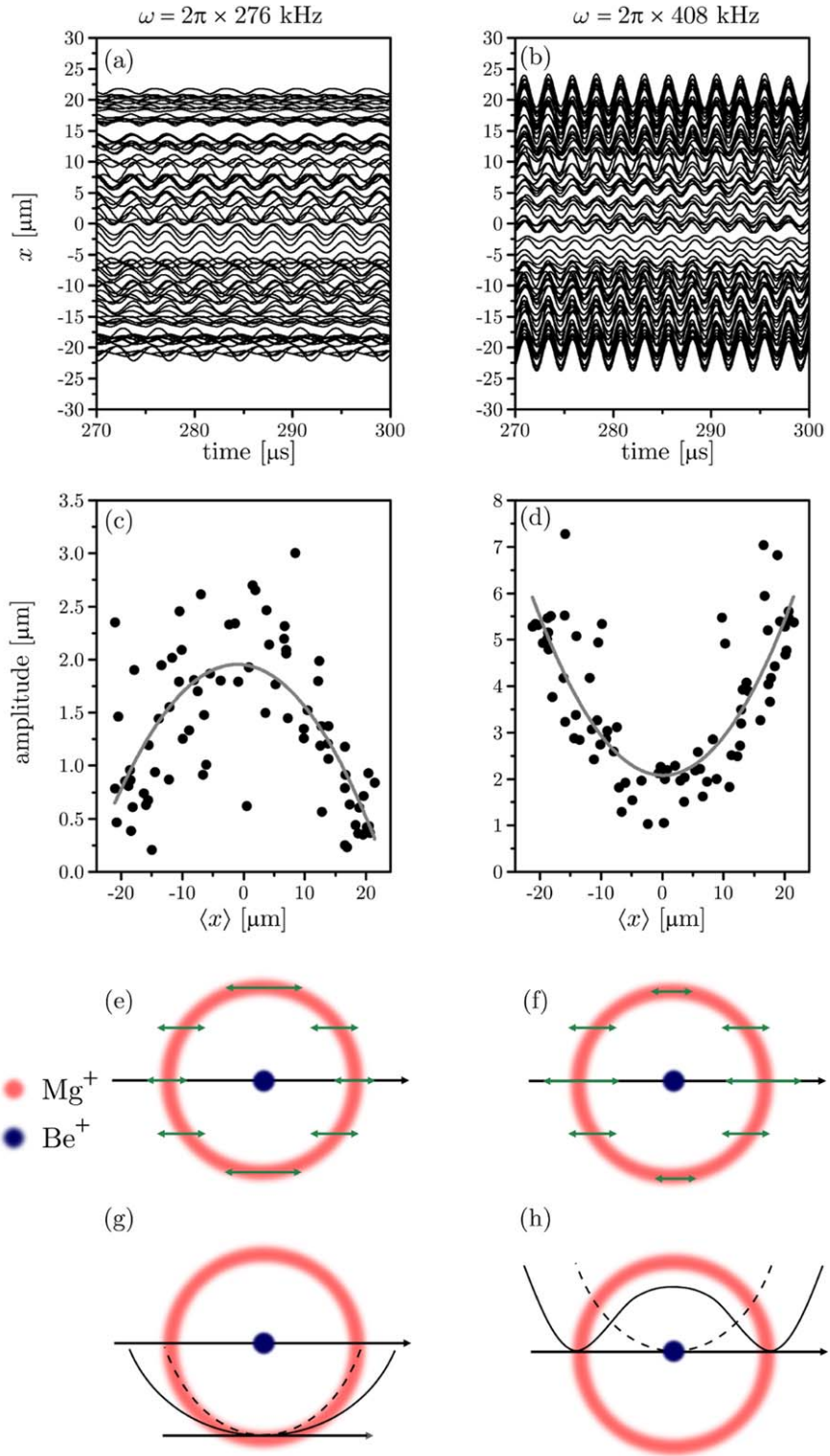

**Figure 4.** (a), (b) Trajectories of the shell ions projected on the $x$-axis when the frequency of the driving force is at either of the components of the split resonance at 276 kHz (left) or 408 kHz (right) (see figure 2(b)). The motional excitation is introduced at $t = 0$, and the oscillatory motions of the ions are found to be stationary at $t > 270 \,\mu s$. (c), (d) Correlation diagrams between the ion amplitudes and time averaged positions $\langle x_i \rangle$. Gray lines in the diagrams are parabolic fits to guide the eye. (e), (f) Axial view of the arrangement of the $Be^+$ core ions (dark blue) and the $Mg^+$ shell ions (light red). The green arrows sketch the motional amplitude of the shell ions (not to scale). The motion also couples to the core ions and makes them oscillate at the same frequencies (not shown in the figure). (g), (h) The trap potential for the shell ions along $x$ direction is depicted with the dashed curves. The solid curves indicate the modified potential due to the interaction with the core ions (not to scale).

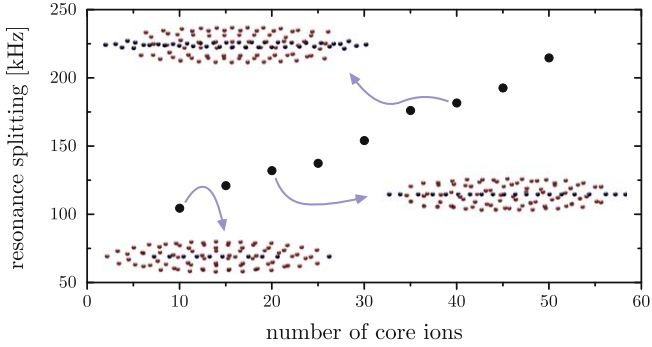

**Figure 5.** The frequency separation between the two split resonances of the shell ions as a function of the number of core ions obtained from molecular dynamics simulations. The number of shell ions is fixed to 80. The insets show the corresponding crystal structure.

For a simplified and intuitive picture, we again invoke the additional potential generated by the other group of ions. Figures 4(g) and (h) show how the core ions (dark blue) modify the potential for the shell ions (light red) along the direction of the excitation field. The dashed curves refer to the trap potential without the core ions, while the solid lines sketch the total potential. As can be seen in figures 4(a) and (b), the shell ions close to $x = 0$ experience a weakening of their potential and hence a decreased resonance frequency, while the ones with larger magnitude of $\langle x_i \rangle$ have their potential enhanced with a corresponding up-shift of their resonance. The situation is, however, less simple than for the modification of the core ions' potential by the shell ions. This has the following two reasons: The effect on ions with intermediate $\langle x_i \rangle$ washes out the splitting of the resonance, which may be represented by the broadening of the down-shifted peak in figure 2(b). In addition, the core ions are not as well fixed in space as the shell ions are when driving the resonance of the core ions around 1 MHz.

Since the modified potential due to the core ions is the reason for the resonance splitting, one expects the splitting to increase with the number of core ions. This is indeed the case as is shown in figure 5.

### 2.5. Conditions for validity of the simple model

The hand-waving arguments for the secular spectra of mixed crystals are valid for large enough numbers of both ion species. The spatial distributions of the ions are not precisely reproducible with every MD simulation run because the initial positions are randomized. However, with a decent number of both ion species, the geometry of the crystal can be always characterized by a nested shell and core structure with radial symmetry. As a result, the features described in section 2.2 are also conserved in its secular spectra (see appendix C for details).

When the number of one species gets much larger than the other, such a simple geometric configuration does not always show up, and the ion distribution is not regularly reproduced for different initial conditions. For the extreme example with only one Mg$^+$ ion within a large number of Be$^+$ ions, the position of the Mg$^+$ ion in the Be$^+$ crystal turns out to be essentially random. The local potential which the Mg$^+$ ion experiences is dependent on its exact position and so is its secular frequency. This is what we see in the experiment and also in the simulations. As a consequence, it is difficult to identify a small number of impurity ions within a larger crystal by their secular frequencies. On the other hand, the majority ions reliably produce secular frequencies close to their single ion resonance.

## 3. Experiments

### 3.1. Experimental setup

The experimental setting is sketched in figure 1(a) with the parameters provided in section 2. It is described in more detail in [34]. The ion crystals are imaged with an objective having a focal length of 76 mm at a distance of about 83 mm from the trap center and are observed with an electron multiplying charge-coupled device camera. Due to chromatic aberration, we cannot simultaneously image the Be$^+$ and Mg$^+$ ions. Instead we move the position of the objective to focus on either of the ion species. An electrical motor stage is employed for this purpose.

To obtain fluorescence from both ion species, we are addressing both cooling transitions, the $2S_{1/2}$–$2P_{3/2}$ (natural linewidth $\gamma/2\pi \approx 19$ MHz) and the $3S_{1/2}$–$3P_{3/2}$ (natural linewidth $\gamma/2\pi \approx 42$ MHz) for Be$^+$ and Mg$^+$ respectively. The cooling lasers at 313 and 280 nm are both detuned by more than 100 MHz. They run axially in opposite directions through the ion trap with intensities of $\approx 300$ W m$^{-2}$ and $\approx 5$ kW m$^{-2}$ for Mg$^+$ and Be$^+$ respectively, determined from the laser powers and beam diameters. We adopted these relatively low intensities in order to reduce distortions of the crystal structure due to radiation pressure.

In addition to the rods forming the quadrupole, there are four outer rod electrodes shown with dashed circles in figure 1(a) (right). Three of these electrodes are utilized to compensate for stray electric fields near the trap center. The excitation force $\vec{f}_{ex}$ is created by applying an ac voltage at frequency $\omega$ to the fourth electrode. Laser cooling is sufficiently effective, such that the ion crystals stay intact when the secular vibrations are excited.

### 3.2. Results and discussion

A mixed Coulomb crystal is presented in figure 6, where the images of Mg$^+$ and Be$^+$ are obtained separately by refocusing the imaging system. Due to its strong chromatic aberration, we observe only a small almost homogeneous background from one ion species when we focus on the other. Therefore, the data was taken with both cooling beams on.

In contrast to our simulations, radiation pressure leads to somewhat asymmetric crystals. This effect could have been avoided by detuning the cooling lasers even further or by reducing their intensity. However, this would require stronger secular excitation due to the reduced fluorescence and hence might eject the ions from the trap. Nonetheless, we believe that it is still reasonable to compare experiment with simulations. By comparing the shape and size of the ion crystal with the simulation, we roughly estimate the number of ions in figure 6 to be 500 for $Mg^+$ and 500 for $Be^+$ [27, 35].

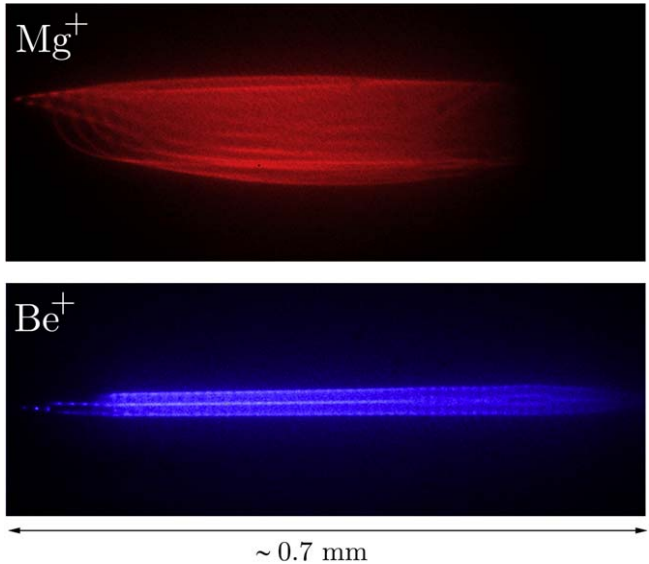

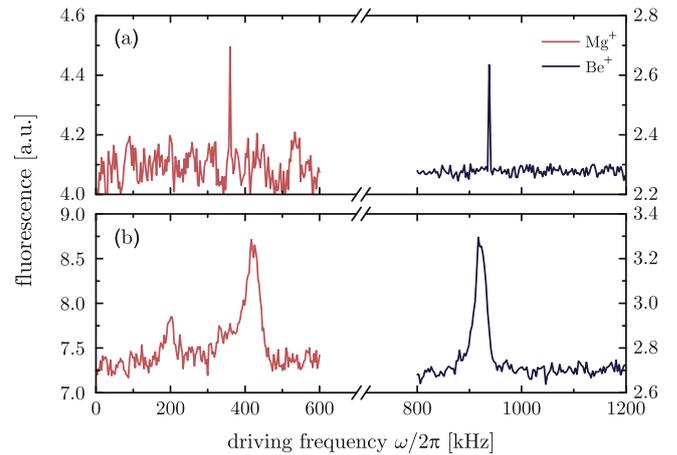

**Figure 6.** Images of a mixed ion crystal comprised of about 500 $Mg^+$ ions and about 500 $Be^+$ ions recorded with an electron multiplying CCD (EMCCD) camera. The fluorescence images of the two ion species are obtained and shown separately, even though they are from the same ion crystal. The asymmetric geometry is attributed to radiation pressure from the cooling beams. Stray potentials and other trap imperfections may also cause such an asymmetry.

**Figure 7.** Experimental secular spectrum of a $Mg^+$ and $Be^+$ ion crystal detected by measuring the fluorescence of the corresponding ion species. (a) Pure $Mg^+$ (light red) and pure $Be^+$ (dark blue) crystals. (b) Two independently prepared mixed crystals, both of which similar to the one presented in figure 6. The light red (left) and dark blue (right) curves are secular spectra detected by observing the fluorescence of the $Mg^+$ and $Be^+$ ions respectively. Therefore, the amplitudes of the left and right parts do not compare and are rescaled. The excitation was done with an amplitude (peak-to-peak) of 0.22 V and 0.24 V for the pure $Mg^+$ and $Be^+$ crystals and with 1.5 V and 1.2 V for the mixed crystals respectively.

Several methods have been demonstrated to estimate the motional amplitude of the ions during secular excitation [34]. In this work we measure the increase of the fluorescence of the ions that is obtained with a largely red detuned cooling laser due to the periodic Doppler shift closer to resonance. The secular spectrum is obtained by scanning the excitation frequency $\omega$. The excitation voltages were carefully optimized to be not too large to disturb the crystal structure, and not too small to observe a sufficient amount of change in fluorescence. They ended up to be between 0.1 and 1.5 V. Based on finite element modeling of our geometry, a voltage modulation of 1 V at one of the outer rod electrodes induces an electric field of 0.75 V m$^{-1}$ at the trap center. This leads to an excitation force that is an order of magnitude stronger than in the simulations. According to our MD simulations, stronger excitation force should mainly influence the cooling time, rather than the secular spectrum of the crystal that is formed. Figure 7 shows an experimental secular motion spectrum that was recorded for a mixed crystal that roughly corresponds in size and composition to the mixed crystal shown in figure 6.

As in the simulations, the resonances of the mixed crystals are significantly broadened in comparison with the pure ion crystals. In addition, we find the spectrum near the pure $Mg^+$ ion crystal resonance to be split into two components, just as in the simulations and the simplified model described above. The observed splitting of 210 kHz is consistent with the simulation performed with a similar number ratio between $Mg^+$ and $Be^+$ ions, considering the uncertainty in the estimation of the number of $Mg^+$ and $Be^+$ ions. We also confirm the broadened resonance near the pure $Be^+$ crystal resonance. The down-shift of this resonance is not predicted by our MD simulations but could be due to a significantly stronger motional excitation in the experiment than in the simulations. The influence of the excitation strength on the secular frequencies has been discussed in [16].

It should be pointed out that the images shown in figure 6 are time averaged and do not necessarily reflect the instantaneous positions of the ions. At finite temperature, ions may exchange their positions from one site to another. We expect the hopping have a minor effect on the secular spectrum shown in figure 7 under our experimental conditions.

## 4. Conclusion

We have investigated the motional resonances of dual-species Coulomb crystals, both theoretically and experimentally. Motional coupling between the two species gives rise to nontrivial characteristics in the motional resonances that do not occur in single-species Coulomb crystals. Both in the MD simulation and in the experiment, we observed that the resonance of the lighter ion species is broadened and shifted to higher frequencies, while the resonance of the heavier species is split into two components. Our findings show that this triple peak structure appears in the secular spectrum of mixed ion crystals even when only two ion species are involved. This could be misinterpreted as an appearance of a third ion species. Our simple model of interactions between core and shell ions qualitatively explains the observation and can help to properly interpret the secular spectrum of mixed ion crystals.

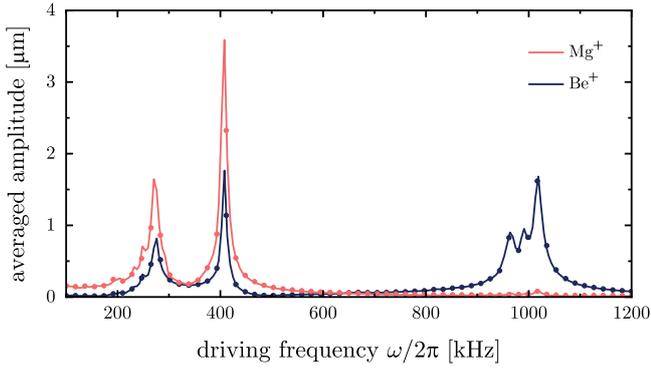

**Figure A1.** A comparison of secular spectra obtained using the time-averaged secular potential (solid lines) and the full time-dependent potential (dots). The secular spectra of the two cases agree well with each other.

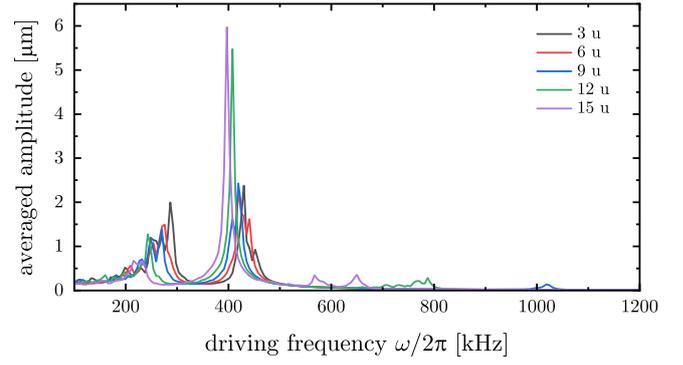

**Figure B1.** Motional spectra of the $^{24}Mg^+$ shell ions for different core ion masses. The number of shell ions is fixed to 80, and the number of core ions to 30.

## Acknowledgments

Byoung-moo Ann acknowledges financial support from DAAD (German academic exchange service) foundation. This project has received funding from the European Research Council (ERC) under the European Union's Horizon 2020 research and innovation programme (grant agreement No. 742247). We also thank Prof Hans A Schuessler for reading the manuscript and giving valuable suggestions.

## Appendix A. Effect of micromotion

Throughout this study, we have utilized the time-averaged effective potential (equation (1)) in the MD simulation rather than the full time-dependent potential of the RF trap. The micromotion induced by the RF potential could affect the space charge distribution of the coolant or the sympathetically cooled ions which may change the mechanical coupling between them. We expect that such an effect is negligible because the time scale of the secular motion is an order of magnitude slower than that of the micromotion. In addition, the amplitude of the micromotion is smaller than the typical ion-to-ion distance. To validate the use of the time-averaged potential, we simulated the secular spectra of a mixed ion crystal including the full time-dependent potential. The result is shown in figure A1. The dots in figure A1 refer to the secular spectra of an ion crystal comprised of 80 $Mg^+$ ions and 20 $Be^+$ ions under the full time-dependent potential, whereas the solid lines refer to the secular spectra given by the same ion crystal, but under the effective potential. The number of samples for the full time-dependent potential case is reduced compared to the effective potential case because of the much longer computation time required for the simulation. The full time-dependent and the effective potential simulation result in almost identical spectra. This confirms that our discussion on the motional coupling is still valid under the influence of micromotion.

## Appendix B. Ion mass dependence

In our simulations we have fixed the mass of the ion species to reflect the experimental conditions. However, we believe that the qualitative features of the motional spectra are quite general and will be reproduced when changing parameters such as ion species, number ratios, and trap depths. As an example, in this section we show how the mass of the lighter (core) ions affects the motional spectrum of the shell ions (section 2.4). We simulated the secular spectra of ion crystals consisting of 80 $^{24}Mg^+$ ions and 30 core ions for core ion masses between 3 and 15 u. The resulting spectra of the $^{24}Mg^+$ ions are shown in figure B1. While there are significant differences between the spectra, the splitting of the single-ion resonance is clearly visible in all of them.

## Appendix C. Randomized ion distribution

When running the MD simulations, the distribution of ions within the resulting mixed crystals depends on the ions' initial positions. Therefore, the precise shapes of the secular spectra are also dependent on the initial conditions. The deviations are more prominent with a weaker cooling force. In this case the resonance widths become narrower so that more details appear in the secular spectrum. Even then, the prominent features in the spectrum discussed in this work are always reproduced. To illustrate this, we have performed 10 MD simulations with 80 $Mg^+$ and 20 $Be^+$ ions under identical trapping conditions, but with different (random) initial conditions. We have used the same parameters given in the main text, except that the driving force as well as the cooling is 10 times weaker in order to enhance the differences. Obtained secular motion spectra are presented in figure C1. Some of the fine structures in the spectra depend on the initial condition as expected. Despite of that, all spectra show a broad and a split resonance and closely resemble figure 2(b).

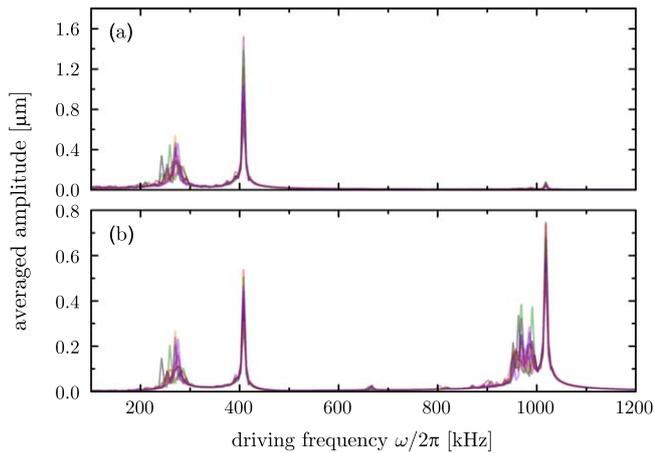

**Figure C1.** 10 secular spectra of a mixed Coulomb crystal comprised of 80 Mg$^+$ ions and 20 Be$^+$ ions obtained with molecular dynamics simulations under identical trapping and cooling conditions, but with different initial conditions. (a) Spectra averaged over all Mg$^+$ ions. (b) Spectra averaged over all Be$^+$ ions. To obtain sharper resonances in order to see finer details, the drag force was reduced to $\alpha = 2.58 \times 10^{-22}$ kg s$^{-1}$ here. Without this, all the spectra of this figure would essentially look identical to figure 2(b).

## ORCID iDs

Fabian Schmid https://orcid.org/0000-0002-9355-180X
Thomas Udem https://orcid.org/0000-0002-9557-5549
Akira Ozawa https://orcid.org/0000-0003-3324-4872